# EMITTANCE GROWTH IN LINEAR INDUCTION ACCELERATORS[*]

**C. A. Ekdahl[ξ], B. T. McCuistian, M. E. Schulze**
*Los Alamos National Laboratory, PO Box 1663*
*Los Alamos, NM 87545, USA*
**C. A. Carlson, D. K. Frayer**
*National Security Technologies, East Gate*
*Los Alamos, NM 87544, USA*
**C. Mostrum, C. H. Thoma**
*Voss Scientific,LLC,*
*Albuquerque, NM , USA*

*Abstract*

The Dual-Axis Radiographic Hydrotest (DARHT) facility uses bremsstrahlung radiation source spots produced by the focused electron beams from two linear induction accelerators (LIAs) to radiograph large hydrodynamic experiments driven by high explosives. Radiographic resolution is determined by the size of the source spot, and beam emittance is the ultimate limitation to spot size. On the DARHT Axis-II LIA we measure an emittance higher than predicted by theoretical simulations, and even though this axis produces sub-millimeter source spots, we are exploring ways to improve the emittance. Some of the possible causes for the discrepancy have been investigated using particle-in-cell (PIC) codes, although most of these are discounted based on beam measurements. The most likely source of emittance growth is a mismatch of the beam to the magnetic transport, which can cause beam halo.

## I. INTRODUCTION

Flash radiography of hydrodynamic experiments driven by high explosives is a well-known diagnostic technique in use at many laboratories [1, 2]. At Los Alamos, the Dual-Axis Radiography for Hydrodynamic Testing (DARHT) facility provides multiple flash radiographs from different directions of an experiment. Two linear induction accelerators (LIAs) make the bremsstrahlung radiographic source spots for orthogonal views. The 2-kA, 20-MeV Axis-I LIA creates a single 60-ns radiography pulse. The 1.7-kA, 16.5-MeV Axis-II creates multiple radiography pulses by kicking them out of a 1600-ns long pulse from the LIA [3-5].

Beam emittance is the ultimate limitation on radiographic source spot size. In the absence of beam-target interaction effects, the spot size is directly proportional to the emittance. Since radiographic resolution is limited by the spot size, minimizing emittance enhances resolution of the radiographs. Therefore, investigation and mitigation of factors leading to high emittance beams would be a productive path to improved radiography.

Improvements in tuning the DARHT Axis-II LIA have reduced the beam motion during the four radiography pulses to less than 1-mm at the accelerator exit [5] and less than 0.04 mm at the final focus [2]. However, the issue of beam emittance has yet to be fully addressed. Although no measurements of the emittance at the diode exit are available, detailed diode simulations with particle-in-cell (PIC) and particle-gun ray-trace codes predict a ~200-300 $\pi$-mm-mr normalized emittance. In the absence of nonlinear forces, the normalized emittance should not vary through the accelerator. However, measurements of emittance in the downstream transport imply an ~800 $\pi$-mm-mr normalized emittance.

Possibilities for this discrepancy are that our modeling of the diode is imperfect, or there is emittance growth in the LIA, or in the transport optics between the LIA and the measurement location. For the present article, we only investigate the possibility of emittance growth in the LIA.

In addition to beam instabilities, there are a number of readily identifiable macroscopic mechanisms for emittance growth in an LIA:

- Solenoid spherical aberrations
- Helical beam trajectories.
- Strong dipole magnetic fields.
- Envelope oscillations.
- Non-uniform current distribution

These mechanisms for emittance growth are amenable to investigation with particle-in-cell (PIC) simulations. A PIC code was used to simulate the beam in the DARHT Axis-II LIA, with particular attention to these mechanisms for emittance growth. The results of these simulations, comparisons with data, and their implications for DARHT are the subject of this article.

---

[*] Research supported by the US Department of Energy and the National Nuclear Security Administration under contract number DE-AC52-06NA253960
[ξ] email: cekdahl@lanl.gov

## II. SIMULATIONS

Two simulation codes were used to explore the causes of emittance growth in a linear induction accelerator: the XTR envelope code and the LSP-slice PIC code. These are described in the next two subsections.

### A. Envelope codes

Design of tunes for the DARHT accelerators is accomplished using envelope codes. The two most frequently used are XTR [6,7] and LAMDA [8]. In both of these codes the radius of a uniform density beam is calculated from an envelope equation [9,10]. In these codes the normalized emittance is taken to be invariant through the accelerator.

The initial conditions for solving the envelope equation are derived from simulations of the diode using the TRAK ray-trace gun design code [3,11] and the LSP PIC code [12].

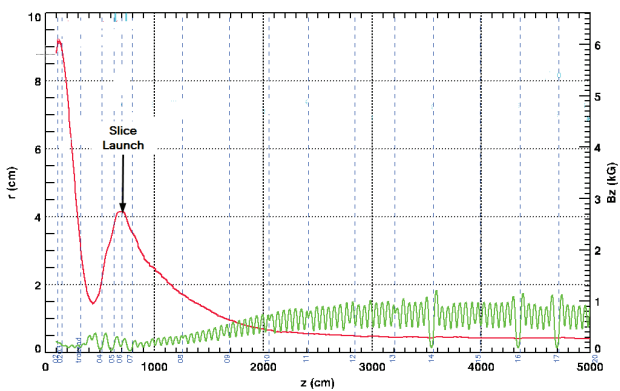

Figure 1: Envelope code simulation of beam transport through the injector cell block and into the main LIA. (Green) The solenoidal focusing magnetic field strength on axis (scale on right). (Red) The beam envelope radius (scale on left). (Blue, Dashed) Locations of beam position monitors (BPMs).

The beam envelope calculated by XTR for the present tune is plotted in Figure 1. The initial envelope focusing is the result of tuning the six injector cells (z<500 cm) to prevent beam spill at any energy in the beam head, which slowly rises from zero to ~2.2 MeV at the diode exit in ~500 ns. The beam then rebounds through a focusing lattice designed to scrape off some of the off-energy beam head. (This region is referred to as the beam-head cleanup zone, or BCUZ.) The beam is then refocused into the main LIA for matched transport with no envelope oscillations.

### B. Particle-in-cell codes

The LSP-slice algorithm is a PIC model for steady-state beam transport based on the Large Scale Plasma (LSP) PIC code [12, 13]. A slice of beam particles located at an incident plane of constant z are initialized on a 2D transverse Cartesian ($x,y$) grid. The use of a Cartesian grid admits non-axisymmetric solutions, including beams that are off axis.

The initial uniform rigid-rotor particle distribution of the slice is extracted from a full $x,y,z$ LSP simulation. The rotation is consistent with zero canonical angular momentum in the given solenoidal magnetic field at the launch position, because the magnetic flux is canceled with a reversed field bucking coil at the DARHT cathode where the beam is created. External fields are input as functions of $z$, and are applied at the instantaneous axial center-of-mass location. LSP-Slice used magnetic fields obtained from XTR. XTR uses axial fields calculated from empirical models of the solenoids, and also includes an algorithm for calculating the transverse dipole fields resulting the measured cell misalignments. External accelerating fields were derived from the locations, geometry, and voltage across the gaps in the XTR simulations.

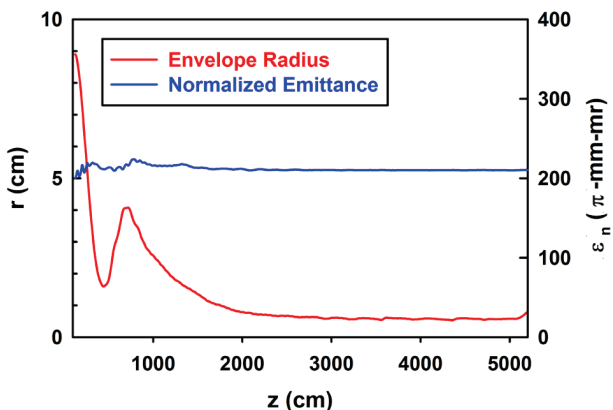

Figure 2: Envelope radius (red curve) and normalized emittance $\varepsilon_n$ (blue curve) calculated by the LSP-Slice PIC code for the tune shown in Fig. 1.

### C. PIC Code results

We based the LSP-Slice simulations on the magnetic tune used on the Axis-II LIA throughout 2013 (Fig. 1). The resulting beam envelope very nearly matched XTR result, and the emittance showed less than ~10 π-mm-mr growth (Fig. 2). The absence of emittance growth for this tune is at odds with experimental estimates of emittance obtained from focusing magnet scans. These show emittance in excess of ~800 π-mm-mr, so it is important to understand the physical root of this discrepancy.

*1) Solenoid spherical aberrations*

Simulations launched at the diode exit showed evidence of edge focusing from the injector solenoids. Edge focusing due to the spherical aberration of solenoids is a well-known effect, especially for large beams [14, 15]. Although little emittance growth resulted (Fig. 2), further simulations were launched at the maxima shown in Fig. 1 to avoid this effect. At this location, the beam envelope is much smaller than at the diode, thereby reducing edge focusing in the simulations.

*1) Offset Injection*

Off center beams can have large helical trajectories in the solenoidal transport field. If the gyro-radius is too large, the beam distribution becomes distorted and the emittance increases. To demonstrate this effect, we initialized the LSP-slice beams with offsets to produce a helical trajectory that encircled the axis. Helical trajectories with gyroradii greater than ~ 1 cm were severely distorted producing latge emittance growth. As measured by our beam position monitors (BPMs) the Axis-II beam is within 1-cm of the axis through the LIA, so emittance growth of more than ~50 π-mm-mr from this mechanism is not expected.

*3) Transverse Magnetic Fields*

Transverse magnetic fields can also produce helical motion. One source of transverse fields in the Axis-II LIA is cell-to-cell misalignment. Although substantial efforts were made to ensure alignment of the magnetic axis, small misalignments exist (~0.025-mm rms offset, and ~0.3-mr rms tilt). Beam energy variations coupling with such misalignments is the source of the "corkscrew" motion [16] observed in other LIAs [17-20]. In DARHT Axis-II this interaction causes a slow beam sweep, which is corrected by application of dipole fields at a few locations in the LIA [5]. The emittance growth caused by the combined misalignment and steering fields is quite small. The LSP-Slice simulations show that these transverse fields produce only a ~10 π-mm-mr emittance increase in the LIA.

*4) Beam Mismatch*

Emittance growth can result from envelope oscillations caused by a mismatch of the beam to the magnetic transport system. A badly mismatched beam exhibits large envelope oscillations, sometimes called a "sausage," "m=0," or "breathing" mode. The detailed mechanism by which this causes emittance growth is parametric amplification of electron orbits that resonate with the envelope oscillation, expelling those electrons from the beam core into a halo [21,22].

Halo growth was quite clear in LSP-slice simulations of mismatched beams [23], and several striking features of this mechanism became evident from the simulation results.

- There was a threshold of oscillation amplitude for noticeable emittance growth.
- When the initial envelope oscillations were small, the emittance grew almost linearly
- When the initial envelope oscillations were large, the emittance rapidly grew and then saturated.
- After the emittance growth saturated, the envelope oscillations were dampened.
- The most severe mismatches showed evidence of multiple halos.

Linear emittance growth on a weakly mismatched beam is illustrated in Fig. 3, and the resulting halo is shown in Fig. 4.

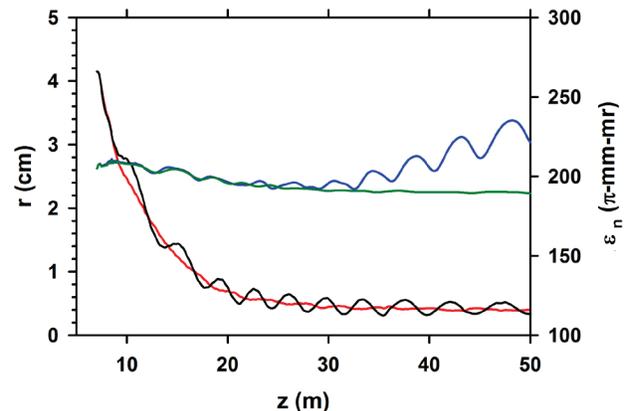

Figure 3: (Black) Envelope radius of a weakly mismatched beam simulated by LSP-Slice. (Red) Envelope radius of a matched beam. (Green) The normalized emittance of the matched beam. (Blue) The normalized emittance of the mismatched beam.

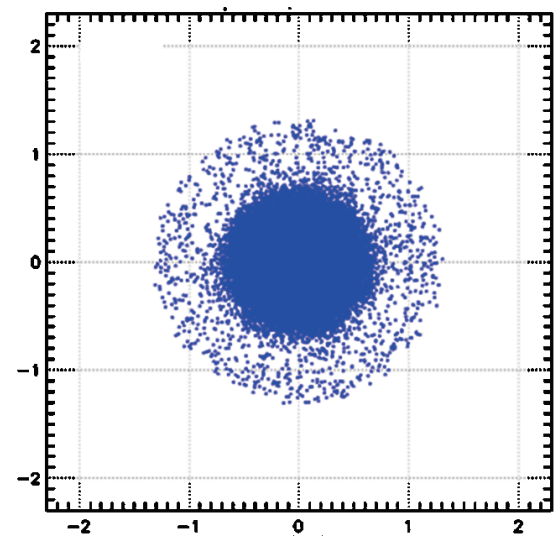

Figure 4: Beam electron distribution at z=52m, which is ~1.5 m past the LIA exit. This is the weakly mismatched case shown in Fig. 3. (This 5-cm x 5-cm Cartesian plot uses a 1-cm grid.)

Rapid emittance growth and saturation on a severely mismatched beam is shown in Fig. 5, and the resulting halo in Fig. 6. The severely mismatched beam emittance grows rapidly and then saturates, apparently by damping of the envelope oscillations as shown in Fig. 5. This effect has also been seen in simulations of ion beams [24]. For this severely mismatched case, the emittance grew to ~900 $\pi$-mm-mr, compared with our measurements of ~800 $\pi$-mm-mr. This suggests that our beam is severely mismatched, if the observed emittance is entirely due to this effect. On the other hand, it is quite possible that other effects in the diode and/or downstream transport also contribute.

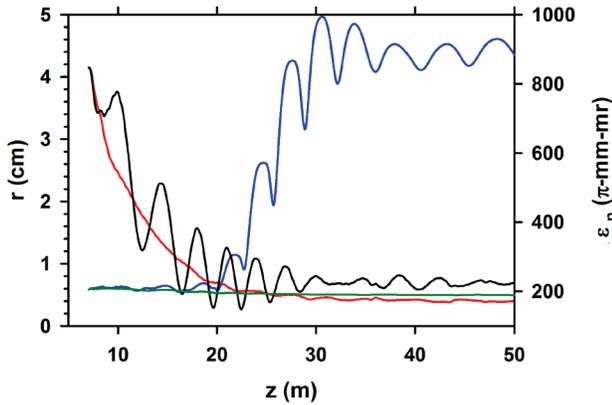

Figure 5: (Black) Envelope radius of a severely mismatched beam showing damping of the oscillations. (Red) Envelope radius of a matched beam. (Green) The normalized emittance of the matched beam. (Blue) The normalized emittance of the mismatched beam, showing rapid growth and saturation. (Note the emittance scale change from Fig. 16).

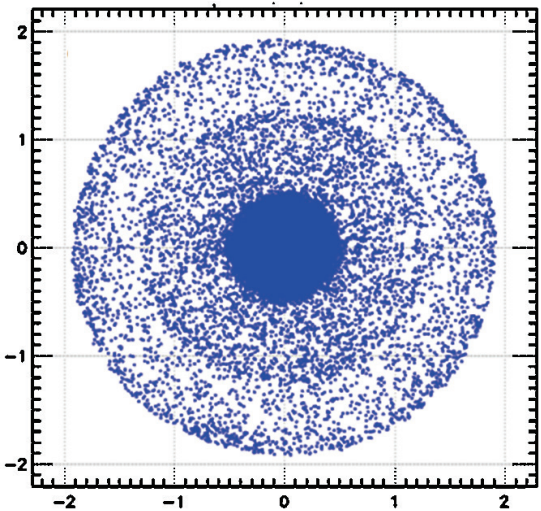

Figure 6: Beam electron distribution at $z = 52$ m, which is ~1.5 m past the LIA exit. This is the severely mismatched case shown in Fig. 5. (This 5-cm x 5-cm Cartesian plot uses a 1-cm grid.)

## III. EXPERIMENTS

We estimated the beam emittance with the focal scan technique, in which a single focusing solenoid is used to vary the beam size at a downstream imaging target. An appropriate beam optics code can then be used to find the beam initial conditions at an upstream point by maximum likelihood fitting to the data [25]. In our measurements we used a solenoid 3.8 m upstream of the final focus to change the size of 50-ns beam pulse produced by the kicker. We imaged the optical transition radiation (OTR) from a 51-micron thick Ti target with a 10-ns gated camera. We used the XTR envelope code to fit our data to find the beam envelope size, divergence, and emittance at a position 3.58 m upstream of the focusing solenoid. The data and XTR fit shown in Fig. 7 yielded 811 $\pi$-mm-mr normalized emittance.

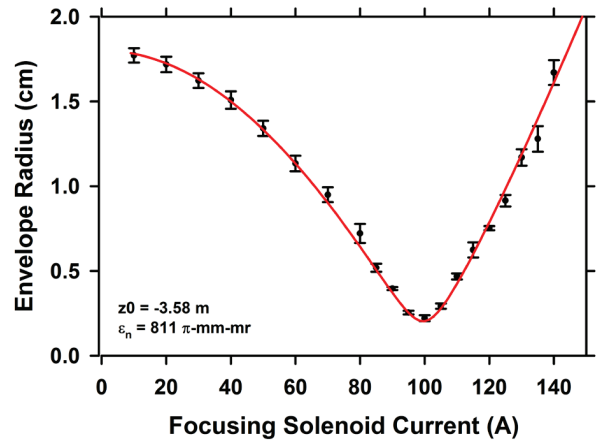

Figure 7: Data from a focal scan of the Axis-2 beam showing the fit by XTR (red line). Error bars indicate uncertainty due to asymmetry of the images.

There are many uncertainties with this technique. Experimentally, the beam may be defocused by ions produced in beam-target interactions or burring of the image by motion from the kicker. These sources of error are partially mitigated by using a short imaging gate. There was a strong radiation produced background, which we corrected by subtraction of a "dark field." (Such background subtractions are an obvious source of uncertainty.) Yet another uncertainty results from the asymmetry of the beam (see Fig 8). To mitigate this we used a beam size calculated from the average of projections of the image into 36 different angles. An overlay of these line spread functions (LSF) for the image in Fig. 8 is shown in Fig. 9.

A final comment about asymmetry is that the envelope theory used to fit the data is itself based on an assumption of an azimuthally symmetric beam. That we use this theory to extract information about an asymmetric beam surely contributes an error.

The beam image (Fig. 8) and the overlay of projections (Fig. 9) show clear evidence of beam halo. However, the halo does not appear to be as wide spread as in the simulation of a severe mismatch giving approximately the same emittance. Figure 10 is the projection of the distribution in Fig. 6, and here it is seen that the halo extends to about 4 beam radii, whereas in the data (Fig. 9) it only extends to about 2 radii. This suggests that much of the emittance observed might be due to other sources.

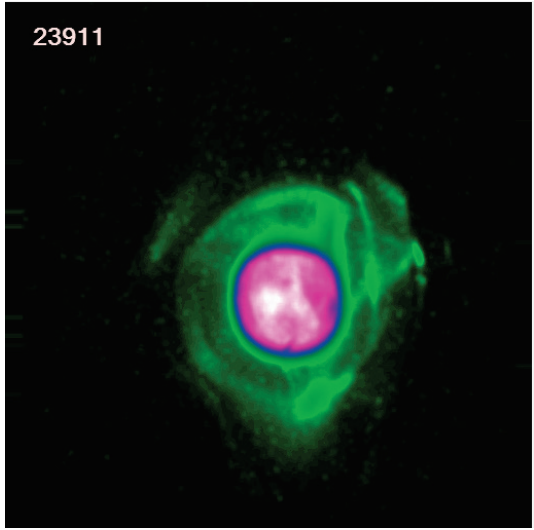

Figure 8: False color image of one of the spots used for the analysis shown in Fig. 7.

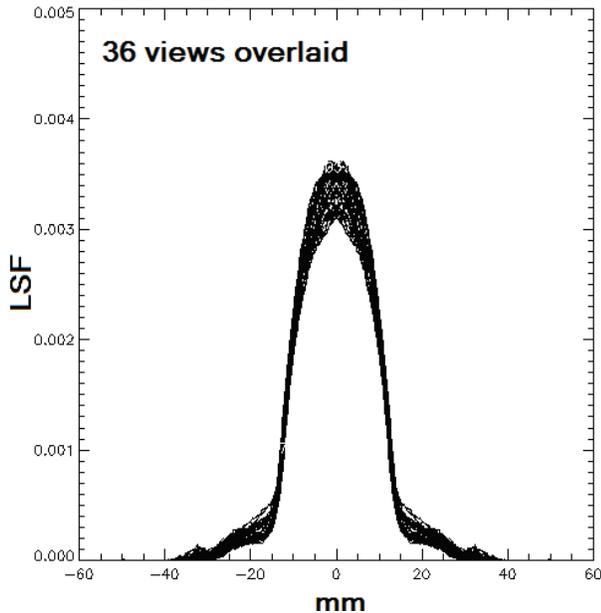

Figure 9: Overlay of line spread functions (LSF) from projections of the image in Fig. 8 into 36 directions differing by 10 degrees.

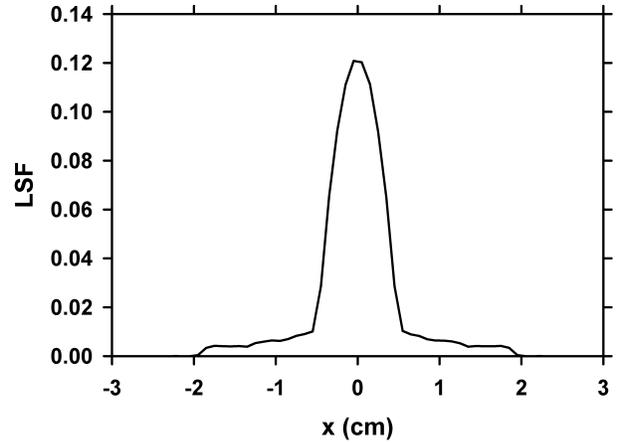

Figure 10: Projection (LSF) of the simulation distribution shown in Fig. 6.

## IV. SUMMARY

The Axis-II beam is centered to less 0.5-cm of the axis at the BPMs throughout most the accelerator, so it is doubtful that there is much emittance growth caused by large gyro-radius effects. Emittance growth from the misalignment dipole fields was shown to be insignificant. Moreover, although we apply a number of steering dipoles to correct for beam motion, these simulations show that growth due to steering dipoles in addition to the misalignments might account for 10 π-mm-mr at most.

If growth in the LIA is indeed responsible for the final ~800 π-mm-mr, suggested by our measurements, then it is most likely due to envelope oscillations resulting from beam mismatch. Halo growth from the parametric amplification of orbits by the envelope oscillations significantly increased the emittance in these simulations. Growth to ~800 π-mm-mr would only result from a very severe mismatch, which would indicate that our diode simulations of initial conditions are grossly inaccurate.

Improving the beam match to reduce envelope oscillations would reduce emittance growth from this cause, and would thereby improve radiographic resolution. The design of our tunes features the ability to improve the match by varying only the first few solenoids after the BCUZ [26]. Therefore, we plan to use this feature in an attempt to reduce the emittance of the DARHT Axis-II beam, in order to improve the radiographic source spot.

As mentioned in the introduction, we will also investigate the possibility that the higher than expected emittance is due to imperfections in the diode, such as non-uniform cathode emission, or inaccurate positioning, which are not accounted for in our simulations to date.